\documentclass[prb,twocolumn,amsmath,amssymb,aps,longbibliography,superscriptaddress]{revtex4-2}

\usepackage{feynmp}
\usepackage{amsmath}
\usepackage{graphicx}
\usepackage{multirow}
\usepackage{latexsym}
\usepackage{textcomp}
\usepackage{verbatim}
\usepackage{color}
\usepackage{bm}
\usepackage{subfigure}
\usepackage{everysel}
\usepackage{keyval}
\usepackage{ragged2e}
\usepackage{dsfont}
\usepackage{amssymb}
\usepackage{enumitem}
\usepackage{pstricks}
\usepackage{mathtools}

\usepackage{braket}
\usepackage{slashed} 
\usepackage{hyperref}        
\hypersetup{
     colorlinks=true,
     citecolor = red,    
     linkcolor=magenta,
     }

\usepackage{graphicx}
\usepackage{xcolor}
\usepackage{amsmath}
\usepackage{bbold}
\usepackage{amsfonts}
\usepackage{bbm}
\usepackage{comment}
\usepackage{orcidlink}

\newcommand{\osum}{{%
    \setbox0\hbox{\circ}%
    \rlap{\hbox to \wd0{\hss\sum\hss}}\box0
}}

\begin{document}

\title{Universal Intrinsic Orbital Dynamics from Berry Curvature in Electronic Two-band Systems}

\author{Jongjun M. Lee\,\orcidlink{0000-0002-9786-1901}}
\thanks{Electronic Address: michaelj.lee@postech.ac.kr}
\affiliation{Department of Physics, Pohang University of Science and Technology (POSTECH), Pohang 37673, Korea}

\begin{abstract}
The geometric structure of quantum states plays a fundamental role in determining the intrinsic dynamics of electrons in solids. In this work, we study the geometric origin of orbital angular momentum and its transport in a general two-band electronic system. Without assuming any symmetry or dimensional constraints, we show that the orbital Berry curvature, which governs the orbital Hall effect, can be universally expressed as the product of the band energy and the square of the Berry curvature. This highlights the central role of Berry curvature in engineering orbital Hall responses. We also discuss the applicability of our framework by analyzing a realistic model. Our findings underscore the geometric universality of itinerant intrinsic orbital dynamics.
\end{abstract}

\date{\today}
\maketitle

\section{Introduction}
The geometric structure of the Bloch wavefunction has been intensively studied in condensed matter physics over the past few decades~\cite{nagaosa2010anomalous,ahn2022riemannian,torma2023essay}. The Berry curvature underlies a wide range of phenomena, including anomalous transport~\cite{haldane2004berry,sodemann2015quantum,liu2018giant}, optical responses~\cite{hosur2011circular,schuler2020local,de2023berry}, magnetoelectric effects~\cite{kato2004observation,lahiri2022nonlinear,liao2024intrinsic}, and light-matter coupling~\cite{mak2018light,lee2023topological}. More broadly, Berry-phase-related geometric quantities provide a unifying framework for understanding various electronic and optical phenomena in solids~\cite{wang2021intrinsic,das2023intrinic,gao2023quantum,wang2023quantum,neupert2013measuring}. While such geometric quantities have been recognized as central to modern condensed matter theory, their precise roles across different physical contexts remain an active area of research~\cite{yu2024non,torma2018quantum,rhim2020quantum,mitscherling2022bound,peotta2015superfluidity,liang2017band,torma2022superconductivity}.

The orbital dynamics of electrons in condensed matter systems have recently attracted attention following experimental observations of associated currents and magnetization~\cite{choi2023observation,el2023observation,lyalin2023magneto,sala2023orbital}. Alongside electron spin, the orbital motion is expected to generate substantial angular momentum and magnetization~\cite{yoda2018orbital,johansson2021spin,chirolli2022colossal,lee2024orbital}, facilitate their transfer~\cite{bernevig2005orbitronics,tanaka2008intrinsic,jo2018gigantic,go2018intrinsic,salemi2022theory}, and lead to slower relaxation processes~\cite{seifert2023time,hayashi2023observation,go2023long,sohn2024dyakonov,moriya2024observation}, marking a new frontier in spintronics~\cite{go2021orbitronics,jo2024spintronics}. However, the physical interpretation of orbital angular momentum and its current remains subtle in periodic lattices~\cite{thonhauser2005orbital,shi2007quantum,pezo2022orbital,pezo2023orbital}. Moreover, although a few proposals have suggested that orbital dynamics are intimately connected to the quantum geometry of electronic states~\cite{pezo2023orbital,piechon2016geometric,urru2025optical}, the connection has not been fully explored through systematic and analytic studies. A geometric framework could help resolve the current gaps in understanding orbital physics in condensed matter systems. Here, we provide such a framework by focusing on the orbital Berry curvature and its geometric origin in generic two-band systems.

\begin{figure}[t!]
    \centering
    \includegraphics[width=0.8\linewidth]{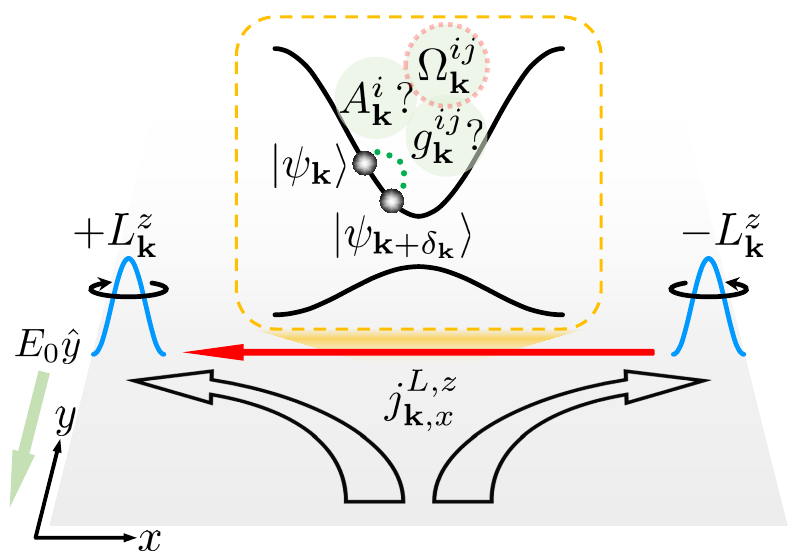}
    \caption{Schematic illustration of the orbital Hall effect and the main result. An electric field $E_0$ applied along the $y$-direction induces a transverse current $j^{L,z}_{\mathbf{k},x}$ along the $x$-direction, carrying orbital angular momentum $\pm L^{z}_{\mathbf{k}}$ polarized along the $z$-direction. This effect originates from the Berry curvature $\Omega^{ij}_{\bf k}$, which characterizes the geometric structure of the electronic wavefunctions.}
    \label{fig1}
\end{figure}

In this study, we provide a direct connection between the Berry curvature and orbital dynamics in general electronic systems. Our analysis does not rely on specific assumptions regarding symmetries or dimensionality, apart from the presence of two well-separated bands. We find that the orbital Berry curvature, which governs the intrinsic orbital Hall effect, can be universally expressed as the product of the band energy and the square of the Berry curvature [Fig.~\ref{fig1}]. This result identifies Berry curvature as the sole geometric quantity responsible for intrinsic orbital transport, and provides a clear route toward engineering orbital Hall responses in real materials. To illustrate the applicability of our framework, we present a quantitative analysis of the Rashba model. Our framework can also be applied to Bogoliubov-de-Gennes Hamiltonians, thereby extending its relevance to anomalous superconductors~\cite{qi2011topological,sigrist1991phenomenological}.

\section{Model and setup}
A general two-band system, non-interacting electrons in arbitrary spatial dimensions, is considered under periodic boundary conditions, without imposing any symmetry constraints. Depending on the system under consideration, the two modes may correspond to different atomic orbitals, spin states, sublattices, or other internal degrees of freedom, and the symmetry analysis may vary accordingly. This system may represent either a genuine two-band model or a low-energy effective theory of a multi-band system in which two bands are well separated from the others~\cite{ashcroft1976solid}. The Hamiltonian density in momentum space is expressed in terms of quadratic fermionic operators as follows.
\begin{equation}
    \mathcal{H}_{\bf k} = \epsilon^{a}_{\bf k}a^{\dagger}_{\bf k}a_{\bf k} +\epsilon^{b}_{\bf k} b^{\dagger}_{\bf k}b_{\bf k} + \gamma_{\bf k} a^{\dagger}_{\bf k}b_{\bf k} + \gamma^{*}_{\bf k} b^{\dagger}_{\bf k}a_{\bf k},
\label{Eq_Ham0_1}
\end{equation}
where $a_{\mathbf{k}}$ and $b_{\mathbf{k}}$ are fermionic annihilation operators with crystal momentum $\mathbf{k}$. The functions $\epsilon^{a}_{\mathbf{k}}$ and $\epsilon^{b}_{\mathbf{k}}$ are real-valued and represent the energy dispersions of each mode. The off-diagonal term $\gamma_{\mathbf{k}} = |\gamma_{\mathbf{k}}| e^{i\theta_{\mathbf{k}}}$ is generally complex, describing the hopping amplitude between the two modes, with a momentum-dependent phase $\theta_{\mathbf{k}}$. Throughout this work, we set $\hbar = 1$ for simplicity and $\epsilon^{a}_{\bf k}>\epsilon^{b}_{\bf k}$ without loss of generality. In matrix form, the Hamiltonian density can be rewritten as follows.
\begin{equation}
    \mathcal{H}_{\bf k} = \Psi^{\dagger}_{\bf k} \mathcal{M}_{\bf k} \Psi_{\bf k}
\end{equation}
where the $2\times 2$ matrix $\mathcal{M}_{\bf k}$ is given by,
\begin{equation}
\mathcal{M}_{\bf k} 
= \begin{pmatrix}
    \epsilon^{a}_{\bf k} & \gamma_{\bf k} \\ \gamma^{*}_{\bf k} & \epsilon^{b}_{\bf k}
\label{Eq_matrix_1}
\end{pmatrix},
\end{equation}
in the basis of the spinor operator $\Psi_{\bf k} = (a_{\bf k},\: b_{\bf k})^{\rm T}$. The eigenvalue problem $\mathcal{M}_{\bf k}|\psi^{\sigma}_{\bf k}\rangle=\epsilon^{\sigma}_{\bf k}|\psi^{\sigma}_{\bf k}\rangle$ diagonalizes the Hamiltonian where $\sigma=\pm$. The eigenstates are given by,
\begin{equation}
    |\psi^{\sigma}_{\bf k}\rangle = \frac{1}{\sqrt{2 \zeta_{\bf k}}}
    \begin{pmatrix}
        \sigma e^{i\theta_{\bf k}} \sqrt{\zeta_{\bf k} +\sigma 1 },& \sqrt{\zeta_{\bf k} +\bar{\sigma}1 }
    \end{pmatrix}^{\rm T},
\end{equation}
and the eigenvalues are given by,
\begin{equation}
    \epsilon^{\sigma}_{\bf k} = \frac{1}{2}\Big( \epsilon^{a}_{\bf k}+\epsilon^{b}_{\bf k} + \sigma \sqrt{(\epsilon^{a}_{\bf k}-\epsilon^{b}_{\bf k})^{2}  + 4|\gamma_{\bf k}|^{2} } \Big),
\end{equation}
where $y_{\bf k} = 2|\gamma_{k}/(\epsilon^{a}_{\bf k}-\epsilon^{b}_{\bf k})|$, $\zeta_{\bf k} = \sqrt{1+y^{2}_{\bf k}}$, and $\bar{\sigma}=\mp$ when $\sigma=\pm$. One can check the eigenstates satisfy the normalization conditions: $\langle \psi^{\sigma}_{\bf k}|\psi^{\sigma'}_{\bf k} \rangle = \delta_{\sigma,\sigma'}$.

\section{Quantum geometric tensor}
The quantum geometric tensor captures the geometric structure of quantum states and is linked to response functions, such as electric conductance~\cite{campos2007quantum,gianfrate2020measurement,kang2025measurements}. In our system, the simple form of the eigenstates enables an analytic evaluation of the quantum geometric tensor. Since the system consists of only two bands, the tensor in the Kubo formula can be expressed in the following form~\cite{ma2010abelian,mera2022nontrivial}.
\begin{equation}
    \eta^{ij,\sigma}_{\bf k} = - \langle \psi^{\sigma}_{\bf k}|\partial_{k_{i}} \psi^{\bar{\sigma}}_{\bf k}\rangle \langle \psi^{\bar{\sigma}}_{\bf k}| \partial_{k_{j}} \psi^{\sigma}_{\bf k}\rangle.
\end{equation}
The quantum geometric tensor is generally complex-valued, with its real part corresponding to the quantum metric $g^{ij,\sigma}_{\mathbf{k}}$ and its imaginary part to the Berry curvature $\Omega^{ij,\sigma}_{\mathbf{k}}$, as given by, $\eta^{ij,\sigma}_{\bf k} = g^{ij,\sigma}_{\bf k} -\frac{i}{2} \Omega^{ij, \sigma}_{\bf k}$~\cite{provost1980riemannian}. Using this formula, we compute the quantum metric and Berry curvature for our two-band model. They take the following simplified exact forms.
\begin{align}
        g^{ij ,\sigma }_{\bf k} &= \frac{1}{4y^{2}_{\bf k}\zeta^{2}_{\bf k}}\Big( \frac{\partial \zeta_{\bf k}}{\partial k_{i}}\frac{\partial \zeta_{\bf k}}{\partial k_{j}}
    +y^{4}_{\bf k} \frac{\partial \theta_{\bf k}}{\partial k_{i}}\frac{\partial \theta_{\bf k}}{\partial k_{j}} \Big), \label{Eq_Metric_1}\\
    \Omega^{ij ,\sigma }_{\bf k} &=  \frac{\sigma}{2\zeta^{2}_{\bf k}} \Big( \frac{\partial \zeta_{\bf k}}{\partial k_{i}}\frac{\partial \theta_{\bf k}}{\partial k_{j}}-\frac{\partial \zeta_{\bf k}}{\partial k_{j}}\frac{\partial \theta_{\bf k}}{\partial k_{i}} \Big).
\label{Eq_Curvature_1}
\end{align}
The quantum metric has the same sign for both bands, whereas the Berry curvature takes opposite signs for the two bands. We also note that the Berry curvature becomes zero if the phase $\theta_{\bf k}$ is a constant in momentum space, in contrast to the quantum metric.

\section{Orbital angular momentum}
We examine the itinerant orbital motion of electrons on a lattice by calculating the orbital angular momentum~\cite{lu2019superconductors,bhowal2021orbital,busch2023orbital,lee2024orbital}. Specifically, we consider the operator $\hat{\mathbf{L}} = \frac{m_{e}}{2} ( \hat{\mathbf{r}} \times \hat{\mathbf{v}} - \hat{\mathbf{v}} \times \hat{\mathbf{r}} )$, which characterizes the angular momentum of delocalized electrons, where $m_{e}$ denotes the electron mass~\cite{pezo2022orbital,pezo2023orbital}. This approach stands in contrast to atom-centered approximations that describe localized orbital motion around atomic nuclei~\cite{go2020orbital,salemi2022first,ding2020harnessing,lee2024composition}.

Recently, a definition of this quantity in periodic lattice systems has been established, avoiding previous conceptual subtleties~\cite{thonhauser2005orbital,shi2007quantum}. Building on these foundations and to ensure gauge invariance, we adopt the following definition of the orbital angular momentum operator~\cite{ceresoli2006orbital,lopez2012wannier,liu2023covariant}.
\begin{equation}
\begin{aligned}
    &\langle \psi^{\sigma}_{\bf k} | \hat{L}^{i}_{\bf k}|\psi^{\sigma'}_{\bf k}\rangle\\
    &= -i\epsilon_{ijk} m_{e}\langle D_{k_{j}}\psi^{\sigma}_{\bf k}| \Big( \mathcal{M}_{\bf k} - \frac{\epsilon^{\sigma}_{\bf k}+\epsilon^{\sigma'}_{\bf k}}{2} \Big) |D_{k_{k}}\psi^{\sigma'}_{\bf k}\rangle,
\end{aligned}
\label{Eq_L_1}
\end{equation}
where $\epsilon_{ijk}$ is the Levi-Civita symbol and $i,j,k = x,y,z$ denote spatial directions. The covariant derivative is defined as $|D_{k_{j}}\psi^{\sigma}_{\bf k}\rangle = |\partial_{k_{j}}\psi^{\sigma}_{\bf k}\rangle + iA^{j,\sigma}_{\bf k}|\psi^{\sigma}_{\bf k}\rangle$, where $A^{j,\sigma}_{\bf k}=i\langle \psi^{\sigma}_{\bf k}|\partial_{k_{j}}\psi^{\sigma}_{\bf k}\rangle$ is the Berry connection. For the diagonal part, the orbital angular momentum for each band is given by,
\begin{equation}
\begin{aligned}
    L^{i,\sigma}_{\bf k} = \epsilon_{ijk} m_{e}  \text{Im} \langle \partial_{k_{j}} \psi^{\sigma}_{\bf k}| (\mathcal{M}_{\bf k}-\epsilon^{\sigma}_{\bf k} )|\partial_{k_{k}} \psi^{\sigma}_{\bf k}\rangle ,
\end{aligned}
\label{Eq_L_2}
\end{equation}
which exactly reproduces the conventional expression for orbital angular momentum without covariant derivatives~\cite{thonhauser2005orbital,shi2007quantum,pezo2022orbital,pezo2023orbital}.

With the eigenstates and eigenvalues in our two-band model, the orbital angular momentum for each band is expressed in the following simple expression.
\begin{equation}
\begin{aligned}
    L^{i,\sigma}_{\bf k} = \sigma \epsilon_{ijk} m_{e} \frac{\epsilon^{+}_{\bf k}-\epsilon^{-}_{\bf k}}{4} \Omega^{jk,\sigma}_{\bf k}.
\label{Eq_OAM_1}
\end{aligned}
\end{equation}
The orbital angular momentum is directly proportional to the Berry curvature [Eq.~(\ref{Eq_Curvature_1})], corresponding to the result in the previous study~\cite{pezo2023orbital}. Both quantities share the same necessary condition for nonzero values. This reveals that the itinerant orbital motion originates from the geometric phase of the Bloch states. The proportionality holds irrespective of the system’s microscopic details, and it remains valid regardless of its symmetry or dimensionality, showing a universal feature.

\section{Transport of the orbital angular momentum}
The orbital Hall effect, characterized by the transverse transport of orbital angular momentum, is governed by the orbital Berry curvature~\cite{go2018intrinsic,bhowal2020intrinsic,phong2023optically,chen2024topology}. We focus on the intrinsic contribution, excluding extrinsic effects from impurity scattering, and the itinerant motion of orbital angular momentum~\cite{gobel2024orbital}. Analogous to the quantum geometric tensor, the orbital Berry curvature is defined through a Kubo formula, as shown below~\cite{pezo2022orbital,pezo2023orbital}.
\begin{equation}
    \mathcal{O}^{ij,\sigma}_{\alpha,\bf k} = -2 \text{Im} \sum_{\sigma'\neq \sigma} \frac{ \langle \psi^{\sigma}_{\bf k}| \hat{j}^{L,\alpha}_{\mathbf{k},i} |\psi^{\sigma'}_{\bf k}\rangle \langle \psi^{\sigma'}_{\bf k}| \partial_{k_{j}}\mathcal{M}_{\bf k}|\psi^{\sigma}_{\bf k}\rangle }{(\epsilon^{\sigma}_{\bf k}-\epsilon^{\sigma'}_{\bf k})^{2}},
\end{equation}
where the orbital current operator is defined by,
\begin{equation}
    \hat{j}^{L,j}_{\mathbf{k},i} = \frac{1}{2}\Big( \hat{L}^{j}_{\bf k} \hat{v}^{i}_{\bf k}  +\hat{v}^{i}_{\bf k} \hat{L}^{j}_{\bf k}  \Big),
\label{Eq_O_Current_1}
\end{equation}
and $\hat{v}^{i}_{\bf k} = \partial_{k_{i}}\mathcal{M}_{\bf k}$. Here, the orbital angular momentum operator $\hat{L}^{j}_{\bf k}$ is defined in Eq.~(\ref{Eq_L_1}). The orbital Berry curvature defined above characterizes the transport of orbital angular momentum in the $\alpha$-direction, associated with motion along the $i$-direction, in response to an applied potential in the $j$-direction.

Based on the above definition, we compute the orbital Berry curvature for our two-band system. The analytical expression is presented below; see the appendix for derivation details.
\begin{equation}
\begin{aligned}
    \mathcal{O}^{ij,\sigma}_{\alpha,\bf k} =&
    - L^{\alpha,\sigma}_{\bf k} \Omega^{ij,\sigma}_{\bf k}.
\label{Eq_OBC_1}
\end{aligned}
\end{equation}
Notably, the orbital Berry curvature is directly proportional to both the orbital angular momentum and the Berry curvature. The orbital Berry curvature does not vanish due to time-reversal symmetry, because both the Berry curvature and the orbital angular momentum are odd under time-reversal operations. Moreover, the orbital angular momentum can be replaced by the Berry curvature using Eq.~(\ref{Eq_OAM_1}). Consequently, the orbital Berry curvature can be expressed entirely in terms of the Berry curvature and the band energy as follows.
\begin{equation}
    \mathcal{O}^{ij,\sigma}_{\alpha,\bf k} =
    - \sigma \epsilon_{\alpha\beta\gamma} m_{e} \frac{\epsilon^{+}_{\bf k}-\epsilon^{-}_{\bf k}}{4}\Omega^{\beta\gamma,\sigma}_{\bf k} \Omega^{ij,\sigma}_{\bf k},
\end{equation}
This expression is universal, as it holds regardless of the system’s microscopic details, symmetries, or dimensionality. Intuitively, this result suggests that one Berry curvature rotates the wavepacket while another drives its transverse motion, collectively producing the orbital Hall effect [Fig.~\ref{fig1}]. The Berry curvature plays a central role in inducing the orbital Hall effect in our two-band system. This constitutes our main result.

\section{Gapped Rashba model}
\begin{figure}[t]
    \centering
    \includegraphics[width=0.98\linewidth]{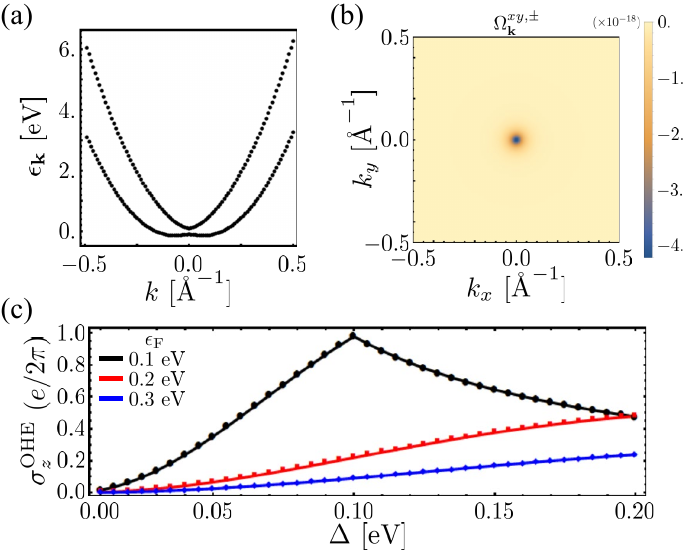}
    \caption{(a) Energy dispersion $\epsilon^{\pm}_{\mathbf{k}}$ of the gapped Rashba model, showing spin-split conduction and valence bands. (b) Berry curvature $\Omega^{xy,+}_{\mathbf{k}}$ of the upper band, exhibiting a monopolar distribution centered at $\mathbf{k} = 0$. (c) Orbital Hall conductivity $\sigma^{\mathrm{OHE}}_z$ as a function of the mass gap $\Delta$ for three different Fermi energies: $\epsilon_{\mathrm{F}} = 0.1\,\text{eV}$, $0.2\,\text{eV}$, and $0.3\,\text{eV}$. The effective mass and spin-orbit coupling are fixed at $m^* = 0.2 m_e$ and $\lambda = 3\,\text{eV}\cdot\text{\AA}$. For (a) and (b), $\Delta = 0.1\,\text{eV}$ is used.}
    \label{fig2}
\end{figure}

To demonstrate the practical applicability of our framework, we consider a two-band model with Rashba-type spin-momentum coupling and a finite mass gap. This corresponds to the parameter choice $\epsilon^{a}_{\bf k} = \frac{k^{2}}{2m^{*}}+\Delta$, $\epsilon^{b}_{\bf k} = \frac{k^{2}}{2m^{*}}-\Delta$, and $\gamma_{\bf k} = -i\lambda(k_{x}-ik_{y})$ in the general two-band Hamiltonian of Eq.~(\ref{Eq_Ham0_1}), where $m^{*}$ is the effective mass, $\lambda$ denotes the strength of the Rashba spin-momentum coupling, and $\Delta$ is the mass gap; see Fig.~\ref{fig2}(a) for the corresponding energy bands. This model captures essential features of two-dimensional systems with broken inversion symmetry and strong spin-orbit coupling. Relevant material platforms include surface alloys such as Bi/Ag(111), oxide heterostructures such as LaAlO$_3$/SrTiO$_3$, and polar semiconductors such as BiTeI, where Rashba splittings on the order of several eV·\AA{} have been experimentally observed~\cite{bihlmayer2007enhanced,caviglia2010tunable,ishizaka2011giant}. The mass gap $\Delta$ can arise from structural asymmetry, substrate-induced effects, or external gating~\cite{shanavas2014electric,krempasky2016entanglement,lin2019interface}.

Using our general expressions, we analytically compute the Berry curvature and orbital Berry curvature. According to Eq.~(\ref{Eq_Curvature_1}), the Berry curvature takes the form
\begin{equation}
\Omega^{xy,\pm}_{\mathbf{k}} = \mp \frac{\Delta \lambda^2}{2( \Delta^2 + \lambda^2 k^2 )^{3/2}},
\end{equation}
where $k^{2}=k^{2}_{x}+k^{2}_{y}$. It is positive or negative definite for each band and exhibits a monopolar structure centered at $\mathbf{k} = 0$, indicating a net anomalous velocity and orbital angular momentum. Figure~\ref{fig2}(b) shows the corresponding momentum-space distribution. According to Eq.~(\ref{Eq_OBC_1}), the orbital Berry curvature is given by the product of the orbital angular momentum and Berry curvature,
\begin{equation}
    \mathcal{O}^{xy,\pm}_{z,\mathbf{k}} = \mp \frac{m_{e}\Delta^{2}\lambda^{4}}{4(\Delta^{2}+\lambda^{2}k^{2})^{5/2}}.
\end{equation}
It follows the same qualitative behavior as the Berry curvature and is sharply peaked near $k = 0$. The dominant contributions to the orbital Hall conductivity originate from these low-momentum regions where geometric effects are strongest.

We then compute the intrinsic orbital Hall conductivity $\sigma^{\mathrm{OHE}}_{z}$ by integrating the orbital Berry curvature over the two-dimensional Brillouin zone.
\begin{equation}
\sigma^{\mathrm{OHE}}_{z} = \sum_{\sigma=\pm} \int \frac{d^2\mathbf{k}}{(2\pi)^2} f(\epsilon^{\sigma}_{\mathbf{k}}) \, \mathcal{O}^{xy,\sigma}_{z,\mathbf{k}},
\end{equation}
where $f(\epsilon^{\sigma}_{\bf k})$ is the Fermi-Dirac distribution and the integration is performed over the occupied states below the Fermi energy $\epsilon_{\rm F}$. Figure~\ref{fig2}(c) shows the resulting conductivity as a function of $\Delta$, evaluated at zero temperature.
The conductivity increases with $\Delta$ until it reaches the Fermi energy and decreases thereafter. 

In our numerical calculations, we have used realistic parameters as follows. $m^* = 0.2 m_e$, $\lambda = 3\,\mathrm{eV}\cdot\text{\AA}$, and $\Delta \le 0.2\,\mathrm{eV}$~\cite{bihlmayer2007enhanced,caviglia2010tunable,ishizaka2011giant}. The orbital Hall conductivity reaches the order of $e/2\pi$, which corresponds to the unit of spin Hall conductance~\cite{bernevig2006quantum,canonico2020two}. Importantly, the orbital Berry curvature retains the same sign regardless of the chirality of the Rashba spin-momentum coupling. This implies that the orbital Hall conductivity scales with the number of Rashba-split bands present in a material. 

\section{Discussion}
Our formalism can be extended to the Bogoliubov-de Gennes Hamiltonian for systems with broken U(1) symmetry due to pair creation and annihilation. In this case, the Hamiltonian density takes a similar form, as shown below.
\begin{equation}
    \mathcal{H}^{\rm BdG}_{\bf k} = \tilde{\Psi}^{\dagger}_{\bf k} \mathcal{M}^{\rm BdG}_{\bf k} \tilde{\Psi}_{\bf k},
\end{equation}
where the $2\times 2$ Hermitian matrix $\mathcal{M}_{\bf k}$ is given by,
\begin{equation}
    \mathcal{M}^{\rm BdG}_{\bf k} = 
    \begin{pmatrix}
        \epsilon_{\bf k} & \gamma_{\bf k} \\
        \gamma^{*}_{\bf k} & -\epsilon_{-\mathbf{k}}
    \end{pmatrix},
\end{equation}
in the Nambu basis $\tilde{\Psi}_{\bf k} = ( a_{\bf k} ,\: a^{\dagger}_{-\mathbf{k}} )^{\rm T}$, and $\gamma_{\bf k} = \gamma_{-\mathbf{k}}$. Since the matrix is similarly diagonalized, the main results can be readily extended by replacing $\epsilon^{a(b)}_{\mathbf{k}}$ with $\epsilon_{+(-)\mathbf{k}}$. Our formalism is directly applicable to two-dimensional anomalous superconductors, such as chiral $p + ip$ states, where the intrinsic orbital response is governed by the geometry of Bogoliubov quasiparticles~\cite{qi2011topological,sigrist1991phenomenological}.

The orbital angular momentum and orbital Berry curvature considered in this work do not account for the localized orbital motion typically described in atom-centered approximations. Instead, they capture only the itinerant orbital dynamics that extend across the system. For instance, it is known that a two-band model based on $p$ orbitals can exhibit the orbital Hall effect in the presence of both inversion and time-reversal symmetries~\cite{han2023microscopic}. In contrast, our result in Eq.~(\ref{Eq_OBC_1}) indicates the absence of the orbital Hall effect in this case, as the imposed symmetries render the phase of the off-diagonal component in the Hamiltonian [Eq.~(\ref{Eq_matrix_1})] momentum-independent. Both itinerant and localized orbital contributions are essential for a complete description of orbital dynamics in periodic lattice systems.

\section{Conclusion and outlook}
In summary, we investigated the orbital angular momentum and its transverse transport in an electronic two-band model on a periodic lattice. No specific constraints, such as symmetry or dimensionality, were assumed in our analysis. We found that the orbital Berry curvature can be universally expressed as the product of the band energy and the square of the Berry curvature. These results clearly demonstrate that intrinsic orbital dynamics originate from the geometric structure of quantum states and highlight the Berry curvature as the central quantity governing orbital transport.

While this study focuses on the orbital Hall effect as a linear response, nonlinear regimes may exhibit even richer geometric structures~\cite{baek2024nonlinear,ovalle2024orbital}. Nonlinear responses may reveal additional geometric contributions beyond the Berry curvature, such as the quantum metric~\cite{ahn2022riemannian,kozii2021intrinsic}. Recent studies further suggest that higher-order geometric structures, such as the quantum Christoffel symbol and quantum Riemann curvature tensor, may play important roles in nonlinear response regimes~\cite{michishita2022dissipation,hetenyi2023fluctuations}. Time-periodic driving, as realized in Floquet systems, may also induce orbital responses that generalize the present results~\cite{oka2019floquet}. In systems with degenerate or nearly-degenerate bands, non-Abelian extensions of the Berry curvature may become essential~\cite{wang2012non,yang2014nonlinear}. Our results further suggest that materials with enhanced Berry curvature, such as polar semiconductors and Rashba systems, offer promising platforms for engineering large orbital Hall effects~\cite{bihlmayer2007enhanced,caviglia2010tunable,ishizaka2011giant}. Finally, while this work has focused on two-band systems, extending the present framework to multi-band models, including possible non-Abelian geometric structures and inter-band coherence effects, remains an important direction for future research~\cite{mitscherling2025orbital}.

\acknowledgements
We thank Jeonghun Sohn, Min Ju Park, and Youngjae Jeon, and especially Hojun Lee, for fruitful discussions. We are also grateful to Hyun-Woo Lee for his continuous support and encouragement throughout the preparation of this study. The data that support the findings of this study are available from the corresponding author upon reasonable request.

\appendix
\section{DETAILED DERIVATION OF THE ORBITAL BERRY CURVATURE}
We begin by substituting the orbital current operator into the Kubo formula for the orbital Berry curvature. This yields the following expression.
\begin{equation}
\begin{aligned}
    \mathcal{O}^{ij,\pm}_{\alpha,\bf k} =& - \frac{1}{(\Delta\epsilon^{\pm}_{\bf k})^{2}} \text{Im}[ \langle \psi^{\pm}_{\bf k}| (\hat{L}^{\alpha}_{\bf k} \partial_{k_{i}}\mathcal{M}_{\bf k} \\
    &+ \partial_{k_{i}}\mathcal{M}_{\bf k} \hat{L}^{\alpha}_{\bf k})  |\psi^{\mp}_{\bf k}\rangle \langle \psi^{\mp}_{\bf k}| \partial_{k_{j}} \mathcal{M}_{\bf k} |\psi^{\pm}_{\bf k}\rangle ],
\end{aligned}
\end{equation}
where $\Delta\epsilon^{\pm}_{\bf k} = \epsilon^{\pm}_{\bf k}-\epsilon^{\mp}_{\bf k}$~\cite{pezo2022orbital,pezo2023orbital}. We then insert the identity operator $\sum_{\sigma}|\psi^{\sigma}_{\bf k}\rangle\langle \psi^{\sigma}_{\bf k}| = \mathcal{I}_{2}$ and rewrite the expression as
\begin{equation}
\begin{aligned}
    \mathcal{O}^{ij,\pm}_{\alpha,\bf k} =& - \frac{1}{(\Delta\epsilon^{\pm}_{\bf k})^{2}} \text{Im}\Big[ \Big(l^{\pm\pm}_{\alpha,\mathbf{k}} M^{\pm\mp}_{i,\mathbf{k}} 
    +l^{\pm\mp}_{\alpha,\mathbf{k}} M^{\mp\mp}_{i,\mathbf{k}}\\
    &+ l^{\pm\mp}_{\alpha,\mathbf{k}} M^{\pm\pm}_{i,\mathbf{k}} 
    +l^{\mp\mp}_{\alpha,\mathbf{k}} M^{\pm\mp}_{i,\mathbf{k}} \Big) M^{\mp\pm}_{j,\mathbf{k}} \Big],
\end{aligned}
\end{equation}
where $l^{\sigma\sigma'}_{\alpha,\mathbf{k}}= \langle \psi^{\sigma}_{\bf k}|\hat{L}^{\alpha}_{\bf k}|\psi^{\sigma'}_{\bf k}\rangle$ and $M^{\sigma\sigma'}_{i,\mathbf{k}} = \langle \psi^{\sigma}_{\bf k}| \partial_{k_{i}}\mathcal{M}_{\bf k}|\psi^{\sigma'}_{\bf k}\rangle$. The orbital angular momentum is given by~\cite{pezo2022orbital,pezo2023orbital}
\begin{equation}
\begin{aligned}
    l^{\sigma\sigma'}_{\alpha,\mathbf{k}} =& \langle \psi^{\sigma}_{\bf k}|\hat{L}^{\alpha}_{\bf k}|\psi^{\sigma'}_{\bf k}\rangle \\
    =& -i \epsilon_{\alpha\beta\gamma}m_{e} \langle D_{k_{\beta}}\psi^{\sigma}_{\bf k}| \Big( \mathcal{M}_{\bf k} - \frac{\epsilon^{\sigma}_{\bf k}+\epsilon^{\sigma'}_{\bf k}}{2} \Big) |D_{k_{\gamma}}\psi^{\sigma'}_{\bf k}\rangle ,
\end{aligned}
\end{equation}
where $|D_{k_{\beta}}\psi^{\sigma}_{\bf k}\rangle= |\partial_{k_{\beta}}\psi^{\sigma}_{\bf k}\rangle+iA^{\beta,\sigma}_{\bf k}|\psi^{\sigma}_{\bf k}\rangle$ is the covariant derivative and $A^{\beta,\sigma}_{\bf k}=i\langle \psi^{\sigma}_{\bf k}|\partial_{k_{\beta}}\psi^{\sigma}_{\bf k}\rangle$ is the Berry connection. The diagonal and off-diagonal terms can be written as
\begin{equation}
\begin{aligned}
    l^{\sigma\sigma}_{\alpha,\mathbf{k}} =& -i \epsilon_{\alpha\beta\gamma}m_{e} \langle \partial_{k_{\beta}}\psi^{\sigma}_{\bf k}| ( \mathcal{M}_{\bf k} - \epsilon^{\sigma}_{\bf k} ) |\partial_{k_{\gamma}}\psi^{\sigma}_{\bf k}\rangle ,\\
    l^{\sigma\bar{\sigma}}_{\alpha,\mathbf{k}} =& -i \epsilon_{\alpha\beta\gamma}m_{e}  \langle \partial_{k_{\beta}}\psi^{\sigma}_{\bf k}| \mathcal{M}_{\bf k} |\partial_{k_{\gamma}}\psi^{\bar{\sigma}}_{\bf k}\rangle \\
    &-\epsilon_{\alpha\beta\gamma}m_{e} \epsilon^{\sigma}_{\bf k} A^{\beta,\sigma}_{\bf k}
    \langle \psi^{\sigma}_{\bf k}|\partial_{k_{\gamma}}\psi^{\bar{\sigma}}_{\bf k}\rangle \\
    &+\epsilon_{\alpha\beta\gamma}m_{e} \epsilon^{\bar{\sigma}}_{\bf k} A^{\gamma,\bar{\sigma}}_{\bf k}
    \langle \partial_{k_{\beta}} \psi^{\sigma}_{\bf k}|\psi^{\bar{\sigma}}_{\bf k}\rangle,
\end{aligned}
\end{equation}
where $\bar{\sigma}=\mp$ if $\sigma=\pm$. We also use the following relations.
\begin{equation}
\begin{aligned}
M^{\pm\mp}_{i,\mathbf{k}} &= -\Delta \epsilon^{\pm}_{\bf k} \langle \psi^{\pm}_{\bf k} | \partial_{k_{i}} \psi^{\mp}_{\bf k} \rangle, \\
M^{\pm\pm}_{i,\mathbf{k}} &= \partial_{k_{i}} \epsilon^{\pm}_{\bf k}.
\end{aligned}
\end{equation}
Noting that the orbital angular momentum has the same sign for both bands and applying the expression of the quantum geometric tensor from the main text, we obtain
\begin{equation}
\begin{aligned}
    \mathcal{O}^{ij,\pm}_{\alpha,\bf k} =& +2 L^{\alpha,\pm}_{\bf k} \text{Im}[\eta^{ij,\pm}_{\bf k}] \\
    &- \frac{\partial_{k_{i}}(\epsilon^{\pm}_{\bf k}+\epsilon^{\mp}_{\bf k})}{\Delta \epsilon^{\pm}_{\bf k}}\text{Im}[l^{\pm\mp}_{\alpha,\mathbf{k}} \langle \psi^{\mp}_{\bf k}|\partial_{k_{j}}\psi^{\pm}_{\bf k}\rangle ] .
\end{aligned}
\end{equation}
The first term on the right-hand side becomes proportional to the Berry curvature. Thus,
\begin{equation}
    \mathcal{O}^{ij,\pm}_{\alpha,\bf k} = \mathcal{O}^{ij,\pm,(1)}_{\alpha,\bf k}+ \mathcal{O}^{ij,\pm,(2)}_{\alpha,\bf k},
\end{equation}
and
\begin{equation}
    \mathcal{O}^{ij,\pm,(1)}_{\alpha,\bf k} = - L^{\alpha,\pm}_{\bf k} \Omega^{ij,\pm}_{\bf k}.
\end{equation}
Using the definition of the orbital angular momentum operator, this second term expression becomes
\begin{equation}
\begin{aligned}
    \mathcal{O}^{ij,\pm,(2)}_{\alpha,\bf k} =&
    - \frac{\partial_{k_{i}}(\epsilon^{\pm}_{\bf k}+\epsilon^{\mp}_{\bf k})}{\Delta \epsilon^{\pm}_{\bf k}} \epsilon_{\alpha\beta\gamma}m_{e}
    \text{Im}\Big[\Big(\\
    &-i \langle \partial_{k_{\beta}}\psi^{\pm}_{\bf k}| \mathcal{M}_{\bf k} |\partial_{k_{\gamma}}\psi^{\mp}_{\bf k}\rangle \\
    &-\epsilon^{\pm}_{\bf k} A^{\beta,\pm}_{\bf k}
    \langle \psi^{\pm}_{\bf k}|\partial_{k_{\gamma}}\psi^{\mp}_{\bf k}\rangle \\
    &-\epsilon^{\mp}_{\bf k} A^{\gamma,\mp}_{\bf k}
    \langle \psi^{\pm}_{\bf k}|\partial_{k_{\beta}} \psi^{\mp}_{\bf k}\rangle
    \Big)
    \langle \psi^{\mp}_{\bf k}|\partial_{k_{j}}\psi^{\pm}_{\bf k}\rangle \Big].
\end{aligned}
\end{equation}
Using the identity, we rewrite the expression.
\begin{equation}
\begin{aligned}
&\langle \partial_{k_{\beta}}\psi^{\pm}_{\bf k}| \mathcal{M}_{\bf k} |\partial_{k_{\gamma}}\psi^{\mp}_{\bf k}\rangle
\langle \psi^{\mp}_{\bf k}|\partial_{k_{j}}\psi^{\pm}_{\bf k}\rangle \\
=& +i \epsilon^{\pm}_{\bf k} A^{\beta,\pm}_{\bf k} \langle \psi^{\pm}_{\bf k} | \partial_{k_{\gamma}} \psi^{\mp}_{\bf k}\rangle \langle \psi^{\mp}_{\bf k}|\partial_{k_{j}}\psi^{\pm}_{\bf k}\rangle  \\
&+i \epsilon^{\mp}_{\bf k} A^{\gamma,\mp}_{\bf k} \langle \psi^{\pm}_{\bf k}| \partial_{k_{\beta}} \psi^{\mp}_{\bf k}\rangle \langle \psi^{\mp}_{\bf k}|\partial_{k_{j}}\psi^{\pm}_{\bf k}\rangle .
\end{aligned}
\end{equation}
The second term of the orbital Berry curvature vanishes.
\begin{equation}
\begin{aligned}
    \mathcal{O}^{ij,\pm,(2)}_{\alpha,\bf k}  =&
    - \frac{\partial_{k_{i}}(\epsilon^{\pm}_{\bf k}+\epsilon^{\mp}_{\bf k})}{\Delta \epsilon^{\pm}_{\bf k}} \epsilon_{\alpha\beta\gamma}m_{e}\\
    &\times \Big(
    +\epsilon^{\pm}_{\bf k} A^{\beta,\pm}_{\bf k} I^{\gamma j,\pm}_{\bf k} 
    +\epsilon^{\mp}_{\bf k} A^{\gamma,\mp}_{\bf k} I^{\beta j,\pm}_{\bf k} \\
    &-\epsilon^{\pm}_{\bf k} A^{\beta,\pm}_{\bf k} I^{\gamma j,\pm}_{\bf k}
    -\epsilon^{\mp}_{\bf k} A^{\gamma,\mp}_{\bf k} I^{\beta j,\pm}_{\bf k}    \Big)\\
    =& 0,
\end{aligned}
\end{equation}
where $I^{ij,\pm}_{\bf k} = \text{Im}[\langle \psi^{\pm}_{\bf k} | \partial_{k_{i}} \psi^{\mp}_{\bf k}\rangle \langle \psi^{\mp}_{\bf k}|\partial_{k_{j}}\psi^{\pm}_{\bf k}\rangle$. Therefore, the orbital Berry curvature becomes
\begin{equation}
\begin{aligned}
    \mathcal{O}^{ij,\pm}_{\alpha,\bf k} =&
    - L^{\alpha,\pm}_{\bf k} \Omega^{ij,\pm}_{\bf k}.
\end{aligned}
\end{equation}
This result corresponds to the expression presented in the main text.

\bibliography{BibRef}

\end{document}